\documentclass[twocolumn,aps,pre,longbibliography]{revtex4-2}

\usepackage{amssymb,amsfonts,amsmath}
\usepackage{euscript}
\usepackage{graphics, graphicx}
\usepackage{setspace}
\usepackage{dcolumn}
\usepackage{mathtools}
\usepackage{bm}
\usepackage{xcolor}
\usepackage[colorlinks=true,citecolor=blue,linkcolor=blue,urlcolor=blue]{hyperref}

\begin{document}

\title{Micro-swimmer collective dynamics in Brinkman flows}

\author{Yasser Almoteri$^{1}$, Enkeleida Lushi$^{2,3}$}
\email[]{YKAlmoteri@imamu.edu.sa, lushi@softactivematter.com}
\affiliation{$^1$ Department of Mathematics and Statistics, Imam Mohammad Ibn Saud Islamic University, Riyadh 11623, Saudi Arabia\\
$^2$ Mathematical Sciences, New Jersey Institute of Technology, Newark, NJ, 07102, United States \\
$^3$ Soft Active Matter Lab, Branchburg, NJ, 08876, United States}

%


\begin{abstract}
Suspensions of swimming micro-organisms are known to undergo intricate collective dynamics as a result of hydrodynamic and collision interactions. Micro-swimmers, such as bacteria and micro-algae, naturally live and have evolved in complex habitats that include impurities, obstacles and interfaces. To elucidate their dynamics in a heterogeneous environment, we consider a continuum theory where the the micro-swimmers are embedded in a Brinkman wet porous medium, which models viscous flow with an additional resistance or friction due to the presence of smaller stationary obstacles. The conservation equation for the swimmer configurations includes advection and rotation by the immersing fluid, and is coupled to the viscous Brinkman fluid flow with an active stress due to the swimmers' motion in it. Resistance alters individual swimmer locomotion and the way it disturbs the surrounding fluid, and thus it alters its hydrodynamic interactions with others and and such affects collective dynamics.The entropy analysis and the linear stability analysis of the system of equations both reveal that resistance delays and hinders the onset and development of the collective swimming instabilities, and can completely suppress it if sufficiently large. Simulations of the full nonlinear system confirm these. We contrast the results with previous theoretical studies on micro-swimmers in homogeneous viscous flow, and discuss relevant experimental realizations. 
\end{abstract}

\maketitle

Micro-organisms are naturally found in many diverse environments, from phytoplankton in the oceans, to microbes in soil and volcanic hot springs, to microbes in plants and animals's bodies  \cite{Lauga09, Pedley92, Ramaswamy10, Marchetti13, Bechinger16, Saintillan18, LiArdekani21, Spagnolie23}. Micro-swimmers such as bacteria are important in nutrient recycling, digestion, fermentation, bioremediation, nitrogen fixation from the atmosphere, and some pathogenic ones can cause infectious diseases. Micro-swimmers such as algae are utilized in photo-bioreactors and drug production. Micro-swimmers such as spermatozoa are crucial in the  propagation of many species of animals. Additionally, artificial active particles such as phoretic colloids or driven colloids \cite{Paxton04, Driscoll17, Karani19} have been inspired by natural micro-swimmers. Given their ubiquity and importance in nature and technology, studying micro-swimmers' behavior, motion, collective dynamics, interactions with the environment, and their response to nutrients or toxins, is paramount to understanding many phenomena they are involved in. 

Many experiments studied how microorganisms or other ``active'' particles affect the fluid they live in, how they interact with each-other through it, and how hydrodynamic and collision interactions affect the emerging collective motion \cite{Dombrowski04, Kim04, Tuval05, Sokolov07, Sokolov09, Drescher11, Mino11, Cisneros11, Sokolov12}. Macroscopic patterns are known to emerge as a result of hydrodynamic interactions and collisions between the micro-swimmers, and they depend on the swimmer type  and geometry \cite{Saintillan07, Saintillan08a, Saintillan08b, Lushi13b, Lushi14, Wioland16}. Significant work has been done on the micro-swimmer collective dynamics emerging due to hydrodynamic or chemotactic interactions or both, e.g. using continuum models \cite{Simha02b, Hill05, Aranson07, Saintillan08a, Baskaran09, Subramanian09, Hohenegger10, Pedley10, Subramanian11, Kasyap12, Lushi12, Ezhilan13, Dunkel13, Kasyap14, Krishnamurthy15, Lushi16, Stenhammar17, Lushi18, Skultety20, Murugan22, Traverso22, Villa-Torrealba23, Almoteri25} or direct particle simulations \cite{Hopkins02, HernandezOrtiz05,  Saintillan07, Ishikawa08, Hernandez-Ortiz09, Saintillan11, Lushi13b, Lushi14, Elgeti15, Wioland16, Zoettl16, Stenhammar17, RojasPerez21}.

Although micro-swimmers naturally live and move in non-trivial environments and confinements such as tissues, soils and sediments, most experimental studies with microbes focus on homogeneous environments such as bulk liquid or flat surfaces \cite{Bechinger16, LiArdekani21, Spagnolie23}. Experimental studies have been conducted on micro-swimmers in complex confinement, for example the colony stability and organization in large drops, racetracks or pillar forests \cite{Wioland13, Lushi14, Wioland16, Nishiguchi18, Makarchuk19, Dehkharghani23}, the motion or transport of swimmers in disordered media with quasi-2D obstacles \cite{Volpe11, Majmudar12, Contino15, Sipos15, Dehkharghani23, Kamdar22}, or in 3D porous environments  \cite{Bhattacharjee19a}. Recent experiments have shown that 3D pore-scale confinement affects micro-swimmer locomotion \cite{Bhattacharjee19a, Bhattacharjee19} and it alters the chemotactic dynamics and morphology of the migrating bacterial population \cite{Bhattacharjee22, MartinezCalvo22, Moore-Ott22}. Theoretical and computational studies of bacterial chemotactic motion in wet porous environments are catching up, e.g. recent work has incorporated the porosity effects by modifying the standard motility parameters from their bulk liquid values \cite{Bhattacharjee22, Alert22}. 

The difficulty with theoretical and computational modeling of active matter in such complex confinements rests with the impossible task of properly and resolving the hydrodynamical interactions as well collisions with the swimmers with each-other and any non-trivially-shaped surface or obstacle in their surrounding. Continuum models are more appropriate to model population dynamics, however one cannot faithfully include the effects of the swimmer collisions, which are crucial in determining the confined dynamics \cite{Wioland13, Lushi14, Wioland16, Nishiguchi18} and can at most approximate them \cite{Ezhilan13}. Numerical simulations of the coupled dynamics of individually-traced micro-swimmers on the other hand are often not feasible for realistic numbers of swimmers because the difficulty and computational cost involved in resolving the non-local interactions such as hydrodynamical ones. Approximations can be made in special cases, e.g. individual swimmers in viscoelastic or power-law flows \cite{Datt15, Datt17, Kos18}, or collective behavior in complex or viscoelastic flows when the intra-swimmer and swimmer-boundary interactions may not be directly included \cite{Bozorgi14, LiArdekani16, Desai17, Stoop19, LiArdekani21, Thijssen21, Kumar22}. An extra special case however is Brinkman flow, which due to its linearity has made possible studies of individual squirmer, helical, or undulatory swimmers moving through it \cite{Leshansky09, Mirbagheri16, Sarah16, Ngangulia18, Nguyen19, Sarah20, Chen20}. 

Despite the relative simplicity of the Brinkman fluid flows, there have not yet been any studies of micro-swimmer collective dynamics in them, which is the question that we address in this paper. Following the coarse-grained models of swimmers in Stokes flow mentioned before, we construct a continuum model that couples the micro-swimmer dynamics to that of the Brinkman viscous flow that they are immersed in.

In the present work we describe a simple continuum model, derived from first principles, to study the dynamics of dilute micro-swimmer suspensions that are immersed in a viscous Brinkman fluid. The model consists of a conservation equation for the swimmer configuration distributions, which is based on the equations for the motion of a self-propelling rod in a local linear flow, and it is coupled to the Brinkman equation for the fluid motion with an extra stress  due to the forcing exerted by the swimmers in it. We analyze the configurational entropy and the stability of the linearized system of uniform isotropic suspensions, and show that the Brinkman resistance hinder the emergence of the long-wave instabilities predicted and well-studied for Stokesian suspensions \cite{Simha02b, Saintillan08a}. We perform simulations of the full nonlinear system where the predictions of the linear analysis are confirmed, and investigate the long-time evolution of of the instabilities and the pattern formation. The analysis and simulations all show that the Brinkman resistance term hinders the onset and development of the collective swimming instabilities.

\vspace{-0.2in}
\section{Swimming in Brinkman flow}
\vspace{-0.2in}

Microorganisms often encounter viscous environments that are heterogeneous and contain stationary obstacles or inert impurities immersed within the fluid medium. Examples are bacteria living in soil \cite{Jung10}, pathogens like bacteria or certain spirochetes navigating heterogeneous tissues in the body \cite{Bhattacharjee19, Bhattacharjee19a}. In such scenarios, to depict the flow of a viscous fluid through spherical particles that are smaller than the flow's characteristic length scale \cite{Cortez10}, one can utilize the {\it Brinkman approximation} \cite{Brinkman47, Brady87}, which can be viewed as incorporating of a lower-order resistance term into the viscous flow equations \cite{Leshansky09, Nguyen19, Chen20}. 

\vspace{-0.2in}
\subsection{Brinkman flow and its disturbances}
\vspace{-0.2in}
The Brinkman Equations to describe viscous flow in a such porous medium are: 
\begin{align}
\mu \nabla^2 \mathbf{u} -\nabla q -(\mu \nu^2)\mathbf{u}= 0, \quad \quad \quad
\nabla \cdot \mathbf{u}= 0
\end{align}
where $\mu$ is the viscosity, $\mathbf{u}$ is the fluid velocity, $q$ is the fluid pressure, and $K_D>0$ is the constant Darcy permeability. Assuming an obstacle length-scale $\ell_p$, 
we obtain 
\begin{align}\label{StoBrinNoFoNonDim}
\nabla^2 \mathbf{u} -\nabla q -\nu^2 \mathbf{u}= 0, \quad \quad \quad
\nabla \cdot \mathbf{u}= 0
\end{align}
where $\nu=\ell_p/ \sqrt K_D$ is the ratio of the particle dimension to the permeability length of the medium \cite{Cortez10, Vanni00}. We will refer to $\nu$ as the resistance parameter. 

The Green's function for the Brinkman Equations, commonly referred to as the {\it Brinkmanlet} \cite{Cortez10} is:
\begin{align}\label{Brinkmanletformula}
\mathbf{B}(\mathbf{r})= H_1(r) \mathbf{I}+ H_2(r) \mathbf{r} \mathbf{r}^T
\end{align}
with $\mathbf{I}$ the identity matrix, $r= | \mathbf{x}-\mathbf{x}_0|$ and  
\begin{align*}
 H_1(r)&= \frac{e^{-\nu r}}{4\pi r}\left(\frac{1}{\nu^2 r^2}+ \frac{1}{\nu r}+1\right)-\frac{1}{4 \pi \nu^2 r^3},\\
 H_2(r)&=  -\frac{e^{-\nu r}}{4\pi r^3}\left(\frac{3}{\nu^2 r^2}+ \frac{3}{\nu r}+1\right)+\frac{3}{4 \pi \nu^2 r^5}.
 \end{align*}
 The fluid flow velocity due to a force $\mathbf{f}$ applied in a Brinkman fluid is then computed by 
 $\mathbf{u}(\mathbf{x})= \mathbf{B}(\mathbf{r})\mathbf{f}$. 
 
\begin{figure}[ht]
\centering
\includegraphics[width=3.4in]{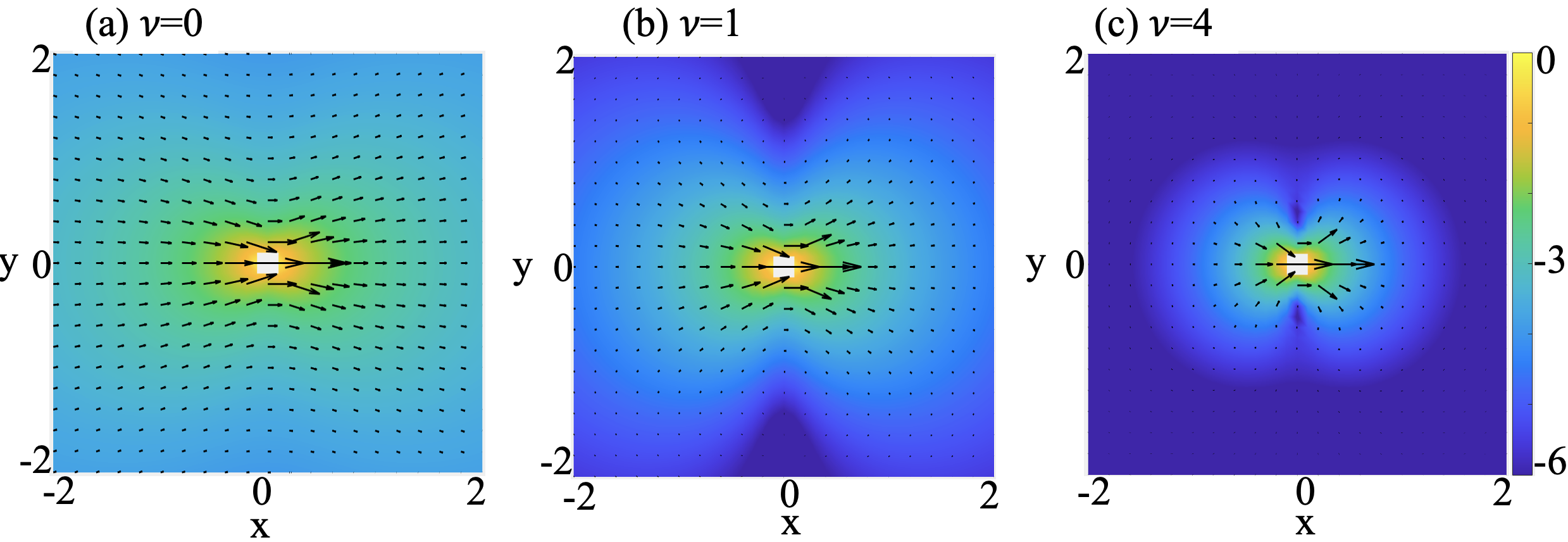}
\vspace{-0.3in}
\caption{Fluid velocity generated when the force $\mathbf{f}=(1,0,0)$ is applied at point $\mathbf{x_0}=(0,0,0)$ in a Brinkman flow with different hydrodynamic resistances: $\nu=0, 1, 4$. The vector field indicates the fluid velocity $\mathbf{u}$, whereas the field color represents $log|\mathbf{u}|$.}
\label{SinStokesAndBrink}
\end{figure}

When a point force is applied to a viscous fluid or Stokes flow, the velocity disturbance that follows it decays as $1/r$ with distance $r$. However in a Brinkman medium the decay of the velocity disturbance is significantly altered at large distances and decays much faster, as seen in Fig. \ref{SinStokesAndBrink}, and at a rate of $1/r^3$ \cite{Brady87} because a new length-scale has been introduced into the problem. 

\vspace{-0.1in}
\vspace{-0.3in}
\begin{figure}[ht]
\centering
\includegraphics[width=3.4in]{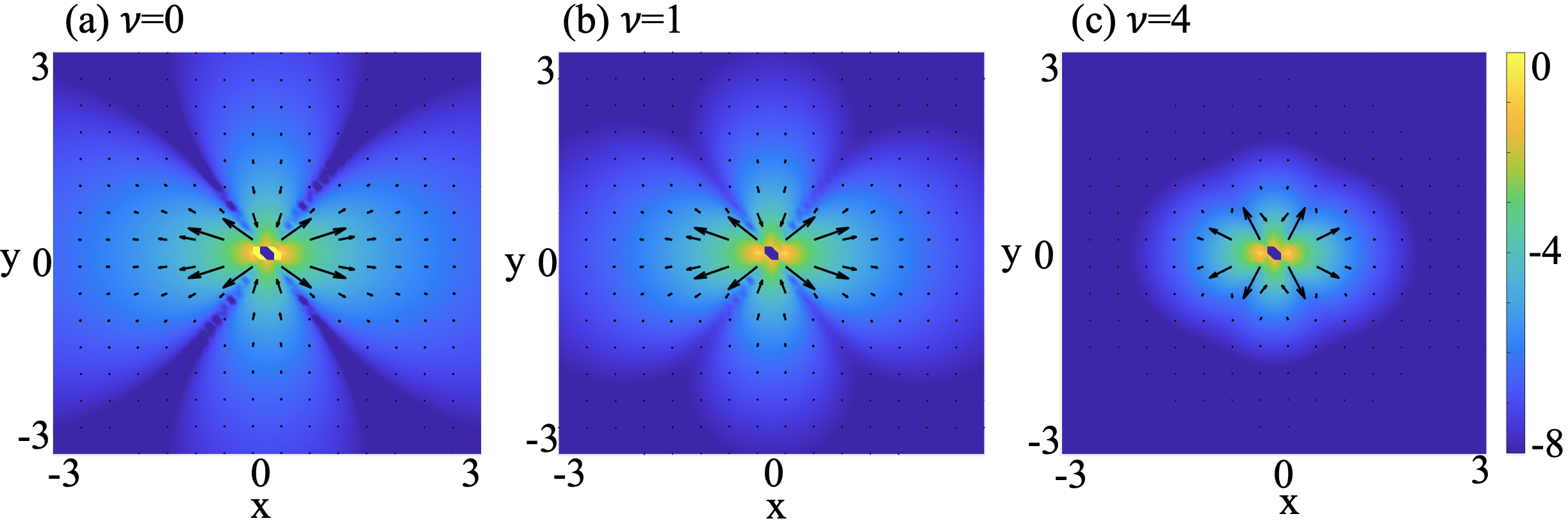}
\vspace{-0.3in}
\caption{Fluid velocity generated by two closely-applied opposite forces in Brinkman flow with different hydrodynamic resistances: $\nu=0, 1, 4$. The vector field indicates the fluid velocity $\mathbf{u}$, whereas the field color represents $log|\mathbf{u}|$.}
\label{SwimmerTwoSinStokesAndBrink}
\end{figure}
\vspace{-0.1in}

In Fig. \ref{SwimmerTwoSinStokesAndBrink} we plot the fluid flow generated by a force dipole (two opposite direction forces applied at a small offset distance) in Brinkman flow of various resistance values. The fluid flow generated by this force dipole, illustrating the highest order of the disturbance fluid flow generated by a micro-swimmer moving through it, is also visibly affected by the medium resistance. In \cite{GuzmanLastra25}, it was shown that for $\nu<1$ the flow was  close to Stokesian and decaying as $1/r^2$ with distance $r$ within a swimmer length, but for $\nu>1$, the flow decays as $1/r^4$ instead.

Figures \ref{SinStokesAndBrink} and \ref{SwimmerTwoSinStokesAndBrink} illustrate the motivation of our study. {\it Given that resistance alters the flow disturbance generated by a swimmer moving through a Brinkman medium \cite{Ngangulia18, Nguyen19, Sarah20}, and given that these fluid disturbances affect the intra-swimmer hydrodynamic interactions that are crucial in the emergence of macroscopic collective behavior in Stokes flow \cite{ Saintillan07, Saintillan08a, Saintillan08b, Lushi13b, Lushi14, Wioland16}, how is the collective swimmer motion altered when they are immersed in Brinkman flow?}

\vspace{-0.2in}
\subsection{Propulsive dumbbell in Brinkman flow}\label{sec:Brinkman_dumbbell_swimmer}
\begin{figure}[ht]
\centering
\includegraphics[width=1.5in]{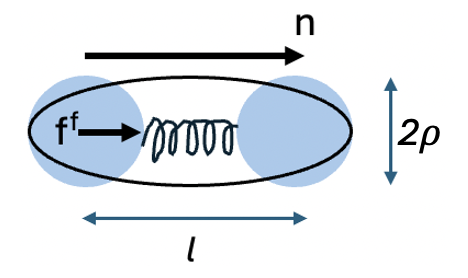}
\vspace{-0.2in}
\caption{Illustration of the dumbbell swimmer.}
\label{DumbbellSwimmer}
\end{figure}

\vspace{-0.2in}
 We first look at the locomotion and hydrodynamics of a single swimmer in Brinkman flow. We adapt the dumbbell model of a Stokesian swimmer by Hernardez-Ortiz et al. \cite{HernandezOrtiz05, Hernandez-Ortiz09}, see Fig. \ref{DumbbellSwimmer}. More details on it and its hydrodynamics can be found in \cite{GuzmanLastra25}. 

The dumbbell swimmer is composed of two equal-sized beads of radius $\rho$, one for the head H with center $ \mathbf{x}_2 $ and the other for the tail T with center $ \mathbf{x}_1$, connected by a FENE spring. A flagellum force $\mathbf{f}^f$ is applied to the tail bead in the direction of $\mathbf{n}= \mathbf{x}_2 - \mathbf{x}_1$, with $\ell = |\mathbf{n}|$, and with strength $f^f$. The swimmer length and width are $\ell+2\rho$ and $2\rho$, respectively. 
\vspace{-0.1in}
\vspace{-0.1in}

On each bead $k=1,2$, $(T,H)$ the force balance is 
\begin{equation}
\mathbf{f}^{h}_{k}+\mathbf{f}^{c}_{k}+ \delta_{k1}\mathbf{f}^{f}_{k}=0.
\end{equation}
Here $\mathbf{f}^{h}$ is the hydrodynamic drag force, $\mathbf{f}^{c}$ is the spring force, and $\mathbf{f}^{f}$ is the flagellum force. 
The hydrodynamic drag force on bead $k$ is given by the drag law for a sphere in Brinkman fluid \cite{Brinkman47, Kim91, Leshansky09}
\begin{equation}
\mathbf{f}^{h}_{k}=\zeta_B (\mathbf{v}_k - \mathbf{u}_k)
\end{equation}
for $\zeta_B=6 \pi \mu \rho (1+\nu + \nu^2/9)$, $\nu$ is the Brinkman resistance parameter (setting $\nu=0$ recovers Stokes drag), $\mathbf{v}_k= d \mathbf{x}_k /dt$ the bead velocity and $\mathbf{u}_k$ the fluid velocity at the bead position. 

These equations give the beads' evolution equations
\begin{equation}
\frac{d\mathbf{x}_k}{dt}= \mathbf{u}_k+ \frac{1}{\zeta_B} (\mathbf{f}^{c}_{k}+ \delta_{1k}\mathbf{f}^{f}_{k}).
\end{equation}
The fluid velocity at some point $\mathbf{x}$ in the domain is 
\begin{align}\label{dumbbellfluidflow}
\mathbf{u}(\mathbf{x})=& \sum_{l=1}^{2} \mathbf{B}(\mathbf{x},\mathbf{x}_l)(- \mathbf{f}^{h}_{l}- \delta_{1l}\mathbf{f}^{f}) = \sum_{l=1}^{2} \mathbf{B}(\mathbf{x},\mathbf{x}_l) \mathbf{f}^{c}_{l}.
\end{align}
Because $\mathbf{B}(\mathbf{x}_1,\mathbf{x}_2)=\mathbf{B}(\mathbf{x}_2,\mathbf{x}_1)$ and $\mathbf{f}^{c}_{1}=-\mathbf{f}^{c}_{2}$, the center of mass $\mathbf{x}_c= (\mathbf{x}_1 + \mathbf{x}_2)/2$ has the dynamics:
\begin{align}\label{xdotoneswimmer}
\frac{d\mathbf{x}_c}{dt} = \frac{1}{2\zeta_B} \mathbf{f}^{f}.
\end{align}
Thus speed for this Brinkman dumbbell swimmer is
\begin{align}
U_B=\frac{U}{1+\nu + \nu^2/9}  := U h(\nu)  \label{Brinkspeeddown}
\end{align}
where $U$ would be the swimmer's speed in Stokes flow and we have named the correction $h(\nu):= 1/(1+\nu + \nu^2/9)$.

Note that Eq. \ref{Brinkspeeddown} is an exact result, and in agreement with work done for the motion of squirmer \cite{Ngangulia18}, helical \cite{Leshansky09, Chen20} and undulatory swimmers \cite{Sarah16, Nguyen19, Sarah20} in Brinkman flows. It indicates that in a fluid laden with sparse tiny obstacles and impurities a single swimmer would move slower (as $\nu>0$) in comparison to a clean viscous fluid (Stokes) where its speed is $U$. 

\vspace{-0.2in}
\subsection{Orientation}\label{sec:Brinkman_dumbbell_swimmer2}
\vspace{-0.2in}

 In the absence of any other swimmers or boundaries, in quiescent Brinkman fluid the swimmer does not change direction, $d\mathbf{n}/dt =0$. 

If in a linear Brinkman fluid, $d\mathbf{n}(\nu)/dt = d\mathbf{n}(\nu=0)/dt +O(\nu^2, \rho^2/\ell^2)$ for $\nu \ll 1$, thus the swimmer orientation is primarily as in Stokes flow with negligible corrections from the resistance. 

\vspace{-0.2in}
\subsection{Extra active stress}\label{sec:Brinkman_dumbbell_swimmer3}
\vspace{-0.2in}

The extra stress on the fluid from the one swimmer is 
\begin{align}
S^{(p)} = \mathbf{f}_1^c \mathbf{x}_1^T + \mathbf{f}_2^c \mathbf{x}_2^T = -\mathbf{f}_1^c  \mathbf{n}^T = - (f^f/\ell) \mathbf{n} \mathbf{n}^T, \label{oneactivestress}
\end{align}
because $\mathbf{n}= \mathbf{x}_2 - \mathbf{x}_1$, $ |\mathbf{n}| =\ell$ , and $\mathbf{f}_1^c = (f^f/\ell) \mathbf{n}$.

Letting $\mathbf{p} = \mathbf{n}/ \ell$ so $|\mathbf{p}|=1$, the dipolar active stress is $S^{(p)} = \sigma_0  \mathbf{p} \mathbf{p}^T$ with $\sigma_0 =-f^f \ell$, and $|\sigma_0|$ is the dipole strength. $\sigma_0<0$ for this bacterium-like pusher swimmer.

Note that the dipole strength $|\sigma_0|$ is the same as in homogenous flow and it not affected by the presence of very sparse obstacles in Brinkman flow. However, the resulting fluid flow from this same-strength dipole, would be altered in a Brinkman fluid, per Eq. \ref{dumbbellfluidflow}.

\vspace{-0.2in}
\section{Mathematical Model}
\vspace{-0.2in}
\subsection{Continuum theory of micro-swimmers in Brinkman flow}
\vspace{-0.2in}

We consider a dilute swimmer suspension in a fluid that also has a very sparse network of stationary obstacles. We assume the intra-obstacle distance as well as the screening length are much larger than the typical swimmer length, thus the resistance parameter is small $\nu \ll 1$.

Coarse-graining and generalizing the equations obtained for the swimmer in Section \ref{sec:Brinkman_dumbbell_swimmer}-\ref{sec:Brinkman_dumbbell_swimmer3}, following standard statistical mechanics approaches \cite{Hohenegger11, Saintillan13, Doi86, Kim91}, we arrive at the probability distribution function $\Psi(\mathbf{x},\mathbf{p},t)$ of the configuration of micro-swimmers, with center of mass $\mathbf{x}$ and orientation $\mathbf{p}$ ($|\mathbf{p}|=1$). The swimmers' dynamics is described by the conservation equation 
\begin{align}\label{ConEq}
 \frac{\partial \Psi}{\partial t} = &- \nabla_x \cdot \left[ \Psi \dot{\mathbf{x}} \right] 
 - \nabla_p \cdot \left[ \Psi \dot{ \mathbf{p}} \right] 
\end{align}
with flux velocities in the center of mass and orientation: 
\begin{align}
\dot{\mathbf{x}} &=  U_B \mathbf{p} + \mathbf{u}- D \nabla_x (\ln \Psi)  \label{xdot}\\
\dot{ \mathbf{p}} &= (\mathbf{I}- \mathbf{p}\mathbf{p}^{ T}) \left[ (\gamma \mathbf{E} + \mathbf{W}) \mathbf{p}\right] - d_r \nabla_{\mathbf{p}}.(\ln{\Psi}). \label{pdot}
\end{align} 
The translational velocity in Eq. (\ref{xdot}) is the sum of the background fluid velocity $\mathbf{u}$ and propulsion with swimming speed $U_B$ along orientation $\mathbf{p}$. Here, as derived before in Eq. \ref{Brinkspeeddown}, $U_B = U/(1+\nu + \nu^2)/9 = U h(\nu)$ where $U$ is the swimmer's speed in Stokes ($\nu=0$) flow.

We include isotropic translational diffusion with a constant coefficient $D$. Here $\nabla_{\mathbf{p}}$ is the gradient operator on the sphere. Eq. (\ref{pdot}) is Jeffery's equation for the angular velocity in terms of the fluid rate-of-strain $\mathbf{E}$, vorticity tensor $\mathbf{W}$ and a shape parameter $-1 \leq \gamma \leq 1$ (where $\gamma=0$ for a sphere and $\gamma \approx 1$ for a rod). It captures the rotation of an anisotropic particle in the local flow. Angular diffusion is included with a constant rotational diffusion coefficient $d_r$ \cite{Saintillan08a, Saintillan08b, Lushi18}. 

Eq. (\ref{ConEq}) does not include a term for micro-swimmer run-and-tumble behavior, though isotropic or anisotropic tumbling is possible to incorporate \cite{Subramanian09, Subramanian11, Lushi12, Lushi16}.

The fluid velocity $\mathbf{u}(\mathbf{x},t)$ satisfies the Brinkman equations with an active stress due to the swimmers motion:
\begin{align}\label{Stokes-Brin}
 -\mu \nabla_x^2 \mathbf{u} +\nabla_x q + \mu/K_D \mathbf{u} = \nabla_x \cdot \Sigma^p, \quad
  \nabla_x \cdot \mathbf{u} = 0.
\end{align}
 $\mu$ is the viscosity, $q$ the fluid pressure, $K_D$ the Darcy permeability constant for the porous medium from Eq. \ref{StoBrinNoFoNonDim}, and $\Sigma^p$ is the active stress due to swimmer, which can be obtained by coarse-graining Eq. (\ref{oneactivestress}):
\begin{align}\label{stress-dim}
 \Sigma^p (\mathbf{x},t) = \sigma_0 \int \Psi (\mathbf{x},\mathbf{p},t) (\mathbf{pp}^T-\mathbf{I}/3)d\mathbf{p}.
\end{align}
$\Sigma^p$ is the average configuration of the active stresses $\sigma_0 (\mathbf{pp}^T-\mathbf{I}/3)$ that are exerted on the fluid by the swimmers over all possible orientations $\mathbf{p}$, as shown in Eq. \ref{oneactivestress} with the term $\mathbf{I}/3)$ added for symmetry \cite{Saintillan08a, Saintillan08b}. $\sigma_0<0$ for {\em pusher} swimmers with rear-activated propulsion, e.g. motile bacteria {\em E. coli}, and  $\sigma_0>0$ for {\em puller} front-activated swimmers, e.g. {\em C. reinhardtii}  \cite{HernandezOrtiz05, Saintillan08a, Saintillan08b, Hernandez-Ortiz09}. 

 We define the local swimmer concentration $\Phi (\mathbf{x},t)$ 
as
\begin{align}
 \Phi(\mathbf{x},t) &= \int \Psi (\mathbf{x},\mathbf{p},t) d \mathbf{p} , \label{PhiEqn} 
\end{align}

As we showed in Section \ref{sec:Brinkman_dumbbell_swimmer}-\ref{sec:Brinkman_dumbbell_swimmer3} that the Brinkman resistance only affects the swimmer speed but not its orientation or the force-dipole strength (to order $O\nu^2$), the only changes in the continuum model for swimmers in the Brinkman fluid here compared to a Stokes fluid \cite{Saintillan08b} lie in Eq. (\ref{xdot}) for the center of mass, namely that $U_b= U/(1+\nu + \nu^2/9)$, and, most importantly, in the presence of the friction term $\nu^2 \mathbf{u}$ in the fluid momentum equations, Eq. \ref{Stokes-Brin}.  Both the $U_B$ correction and the friction term include the influence of the very sparse obstacle network onto the swimmers as a bulk fluid effect instead of through more direct interactions. 

\vspace{-0.2in}
\subsection{Non-dimensionalization of the system}\label{nondimsection}
\vspace{-0.2in}

We non-dimensionalize  Eq. (\ref{ConEq}) using the distribution, velocity, length and time scales $ \Psi_c=n$, $u_c=U$, $l_c=(nl^2)^{-1}$, $t_c = l_c/u_c$, with $l_c=(V/V_p)l$ and $V_p=Nl^3$ the effective volume taken by $N$ swimmers of length $l$ in the fluid volume $V$ of a cube with length $L$ \cite{Saintillan08a}. $n$ is the mean number density of the micro-swimmers in the volume $V$. This choice of non-dimensionalization normalizes the distribution function  
$(1/V) \int_V d\mathbf{x} \int  \Psi(\mathbf{x},\mathbf{p},t) d\mathbf{p}= 1$
with $\Psi_0= 1/4\pi$ the probability density for the uniform isotropic state.

We also non-dimensionalize the diffusions as 
$D' = D t_c/l^2_c$, $d'_r = d_r t_c$. Dropping the $'$-s, the non-dimensionalized swimmer distribution equation becomes:
\begin{align}\label{PsiEqnNondimensional}
 \frac{\partial \Psi}{\partial t} = &- \nabla_x \cdot [ \Psi (h(\nu) \mathbf{p} + \mathbf{u} ) ] \nonumber \\ 
 &- \nabla_p \cdot [ \Psi  (\mathbf{I}-\mathbf{pp}^T ) (\gamma \mathbf{E} + \mathbf{W}) \mathbf{p} ] \nonumber \\ 
  &+D \nabla_x^2 \Psi + d_r \nabla_p^2 \Psi. 
\end{align}
The fluid equations are non-dimensionalized using $\alpha= \sigma_0 / ( U \mu l_c^2 )$ for $\alpha$ a non-dimensional $O(1)$ constant whose sign tells whether the micro-swimmers are pushers ($\alpha<0$), pullers ($\alpha>0$)\cite{Saintillan08a, Lushi18}. The non-dimensional resistance parameter is $\nu=  l_c/\sqrt{K_D}$, and we will refer to is as the hydrodynamic resistance (note the difference from Eq. (\ref{StoBrinNoFoNonDim}) ). The non-dimensional Brinkman fluid equations with an active stress are:
\begin{align}\label{nondimfluid}
 &-\nabla_x^2 \mathbf{u} + \nabla_x q + \nu^2 \mathbf{u } = \nabla_x \cdot  \Sigma^p, \quad \quad \nabla_x \cdot \mathbf{u} = 0  \nonumber \\
 &\Sigma^p =\alpha  \int \Psi (\mathbf{x},\mathbf{p},t) [\mathbf{pp}^T-\mathbf{I}/3]d\mathbf{p}.
\end{align}

We make a note on the unusual non-dimensionalizaton choice for the characteristic lengthscale  $l_c=(V/V_p)l=1/(nl^2)$, which largely follows previous work on this topic \cite{Saintillan08a, Saintillan08b, Lushi12, Ezhilan13, Lushi18}. See that $\alpha = (\sigma_0/\mu U_0 l^2)N (l/L)^3 $ and $\nu = (1/K_D)N (\ell/L)^3$, so the number of swimmers $N$ and hence their mean number density $n$ is encapsulated into the strength of the swimmer effect into the fluid, $\alpha$, and well as the strength of the resistance, $\nu$.

\vspace{-0.2in}
\section{Configurational Entropy}\label{sectionEntropy}
\vspace{-0.2in}
As in work for Stokesian swimmer suspensions \cite{Saintillan08b, Hohenegger10}, we define the system's configurational entropy $\mathcal{S}$ as
\begin{align}\label{EnEq}
\mathcal{S} = \int s  d\mathbf{x} =  \int \int  (\Psi /\Psi_0) \ln(\Psi /\Psi_0) d\mathbf{p} d\mathbf{x},
\end{align}
with $\Psi_0 = 1/4\pi$ and  $s= \int (\Psi/\Psi_0) \ln(\Psi/\Psi_0) d\mathbf{p} $. Note that $\mathcal{S}= \int  s(\mathbf{x},t) d\mathbf{x} \geq 0$ and $\mathcal{S}$ realizes its minimum value of zero at $\Psi_0$, the homogeneous isotropic state.

Differentiating $s(\mathbf{x},t)$ in time and using the identity $ \nabla_p  \cdot [(\mathbf{I}-\mathbf{p}\mathbf{p}^{ T}) \nabla_x \mathbf{u} \, \mathbf{p}] = -3 \mathbf{p}\mathbf{p}^{ T} : \mathbf{E}$, we obtain: 
\begin{align}\label{smallseq}
&s_t+ \mathbf{u} \cdot \nabla_x s +\nabla_x \cdot \int U_B \mathbf{p} (\Psi/\Psi_0) \ln (\Psi/\Psi_0)  d\mathbf{p}  \nonumber \\
&= D \nabla^{2}_x s + 3 \left( \int  (\Psi/\Psi_0) \mathbf{p}\mathbf{p}^{ T} d\mathbf{p} \right) : \mathbf{E}  \nonumber \\
&-\int \left[ D |\nabla_x \ln (\Psi/\Psi_0) |^2 + d_r |\nabla_p \ln (\Psi/\Psi_0) |^2 \right] d\mathbf{p}. 
\end{align}
The momentum equation in Eq. (\ref{nondimfluid}) can be integrated over the fluid domain to get:
\begin{align}\label{EnEq1}
 &2 \int   \mathbf{E} :  \mathbf{E} d\mathbf{x} + \nu^2 \int  |\mathbf{u}|^2 d\mathbf{x} = - \int  \mathbf{E} : \Sigma^p d\mathbf{x} .  \nonumber \\
 &= -\alpha \Psi_0  \int  \mathbf{E} : \left( \int  (\Psi/\Psi_0) \mathbf{p}\mathbf{p}^{ T} d\mathbf{p} \right) d\mathbf{x}.
\end{align}
Integrating Eq. (\ref{smallseq}) in $\mathbf{x}$ and using the Eq. (\ref{EnEq1}) gives, ultimately gives an {\em exact} equation for the evolution of the configurational entropy:
\begin{align}\label{EnthroEqua}
&\Psi_0 \mathcal{D}_t S = \frac{-6}{\alpha} \int \mathbf{E} :\mathbf{E} d\mathbf{x} - 3\nu^2 \int |\mathbf{u}|^2 d\mathbf{x} \nonumber \\
&- \int  \int \Psi [D |\nabla_x \ln \Psi |^2 + d_r |\nabla_p \ln \Psi |^2] d\mathbf{p}  d\mathbf{x}. 
\end{align}
In Eq. (\ref{EnthroEqua}), we notice three distinctive contributions to the configurational entropy from the hydrodynamics, resistance, and diffusive processes. 

The first term in the right hand side of Eq. (\ref{EnthroEqua}) contains the rate of viscous dissipation $\int d\mathbf{x}  \mathbf{E} :\mathbf{E}$ which is positive definite, hence, in the absence of any external forcing or boundaries, for suspensions of pullers ($\alpha>0$) any fluctuations from the isotropic state as measured by the entropy are expected to monotonically dissipate, whereas for suspensions of pushers ($\alpha<0$) there is a feedback loop where fluctuations create velocity gradients which further increase fluctuations \cite{Saintillan08a, Saintillan08b}. These are balanced by the diffusive processes in the system, seen in the third term in the right hand side of Eq. (\ref{EnthroEqua}). 

The hydrodynamic resistance makes an appearance in the second term of Eq. (\ref{EnthroEqua}). As $\int d\mathbf{x} |\mathbf{u}|^2$ is positive definite and $\nu^2 \geq 0$, it is clear that the resistance is expected to dampen any fluctuations from the isotropic state, thus having a stabilizing effect on the system. This expectation is supported by the results of our linear stability analysis of the isotropic state and by the results of our nonlinear simulations, presented in subsequent sections. 

Notable in Eq. (\ref{EnthroEqua}) is the absence of the advection term $\nabla_x \cdot \int U_B \mathbf{p} (\Psi/\Psi_0) \ln (\Psi/\Psi_0)  d\mathbf{p} $ seen in the evolution of $s$ in Eq.(\ref{smallseq}) because it disappears in the $\mathbf{x}$-integration. This indicates that {\em the Brinkman resistance primarily affects the system dynamics through the bulk-fluid effect on the suspension and not through the decrease in the individual swimmer speed in $U_B$.}

\vspace{-0.2in}
\section{Linear Stability}
\vspace{-0.1in}
\subsection{The eigenvalue problem} \label{TheEigenvalueProblem}
\vspace{-0.1in}

We consider the stability of a nearly uniform and isotropic swimmer suspension. For simplicity, we neglect angular diffusion $d_r=0$ but  retain translational diffusion $D$. Let the swimmer suspension be a plane wave perturbation about the uniform isotropic state $\Psi = \Psi_0 =1/4\pi$ and zero fluid flow $\mathbf{u_0}=\mathbf{0}$ as:
\begin{align}
\Psi(\mathbf{x}, \mathbf{p},t) &= 1/4\pi + \epsilon \tilde{\Psi}(\mathbf{p},\mathbf{k}) \exp (i \mathbf{k} \cdot \mathbf{x} + \sigma t) \nonumber \\
\mathbf{u}(\mathbf{x}, \mathbf{p},t) &=\mathbf{0} + \epsilon \tilde{\mathbf{u}}(\mathbf{p},\mathbf{k}) \exp (i \mathbf{k} \cdot \mathbf{x} + \sigma t),   \nonumber%
\end{align}
with $|\epsilon| \ll1$, $\mathbf{k}$ the wave-numbers, $\sigma(k)$ the growth rate. 

Substituting these into the linearized system, we get an equation that is linear in $\tilde{\Psi}$: 
\begin{align}
(\sigma + i h(\nu) \mathbf{p} \cdot \mathbf{k} + D k^2) \tilde{\Psi}&= \frac{3}{2} i \gamma \Psi_0 \mathbf{p^T}(\tilde{\mathbf{u}}\mathbf{k^T} + \mathbf{k} \tilde{\mathbf{u}}^T)\mathbf{p}.  \nonumber
\end{align}
Letting $\mathbf{k}=k\hat{\mathbf{k}}$, we solve the fluid equations 
\begin{align}
\tilde{\mathbf{u}} = \frac{i k}{k^2 +\nu^2}(\mathbf{I} - \hat{\mathbf{k}}\hat{\mathbf{k}}^T)  \tilde{\Sigma^p}  \hat{\mathbf{k}}, \label{teldaveloci} \\ 
\tilde{\Sigma^p} = \alpha  \int \tilde{\Psi'} \mathbf{p'p'}^T d\mathbf{p'}. 
\end{align}
Finally, plugging this into the $\tilde{\Psi}$ equation, we get
\begin{align}\label{psi111}
&\left(\sigma + i h(\nu) \mathbf{p} \cdot \mathbf{k} + D k^2 \right) \tilde{\Psi} \nonumber \\
= &- \frac{3\gamma k^2}{4\pi (k^2+\nu^2)} \mathbf{p}^T (\mathbf{I}-\hat{\mathbf{k}}\hat{\mathbf{k}}^T) \tilde{\Sigma}^p \hat{\mathbf{k}}\hat{\mathbf{k}}^T \mathbf{p}.
\end{align} 

Without loss of generality, we let $\mathbf{\hat{k}} = \mathbf{\hat{z}}$, $\mathbf{p}=\left[ \sin \theta \cos \phi, \sin \theta \sin \phi, \cos \theta \right]$ and $d \mathbf{p} = \sin \theta d\theta d\phi$ for $\theta \in [0, \pi]$ , $\phi \in [0, 2\pi)$. Then, we can write 
\begin{align}
&(\sigma+ik h(\nu) \cos \theta+ D k^2) \tilde{\Psi} \nonumber \\
= &- \frac{3\gamma \alpha k^2}{4\pi (k^2+\nu^2)} \cos \theta \sin \theta [ \cos \phi F_1 + \sin \phi F_2], 
\end{align}
where we have defined the integral operators of $\tilde{\Psi}$
\begin{align}
F_1(\tilde{\Psi})&= \int_0^{2\pi}  \cos \phi' \int_0^{\pi} \sin^2 \theta' \cos \theta' \tilde{\Psi}(\theta', \phi') d\theta' d\phi'  \nonumber \\
F_2 (\tilde{\Psi})&=\int_0^{2\pi}  \sin \phi' \int_0^{\pi} \sin^2 \theta' \cos \theta' \tilde{\Psi}(\theta', \phi') d\theta' d\phi'  \nonumber 
\end{align}

Eq. (\ref{psi111}) constitutes a linear eigenvalue problem for the perturbation mode $\tilde{\Psi}$ and the growth rate $\sigma$. Applying  the operator $F_1$ to Eq. (\ref{psi111}), we arrive at an integral equation for $\sigma$:
 \begin{align}\label{sigmainteqn}
1&=  \frac{-3\gamma \alpha k^2}{4\pi (k^2+\nu^2)} \int_0^{\pi} \frac{\sin^3 \theta \cos^2 \theta}{(\sigma+ik h(\nu) \cos \theta+ D k^2)} d\theta. 
\end{align}

With $a=(\sigma + D k^2)/(ik h(\nu))$, we can evaluate this to:
\begin{align}
1 = \frac{-3 \alpha \gamma k^2}{4(k^2 + \nu^2)} \frac{1}{ik} \left[  2a^3 -\frac{4a}{3} +(a^4-a^2)\log \left( \frac{a-1}{a+1} \right)  \right], \nonumber
\end{align}
which we rewrite as
\begin{align} \label{Hrelation}
&0 =\mathcal{F}(\sigma,k) := \nonumber \\
&\frac{-4ik}{3(-\alpha \gamma)} \frac{k^2 + \nu^2}{k^2}+\left[  2a^3 -\frac{4a}{3} +(a^4-a^2)\log \left( \frac{a-1}{a+1} \right)  \right].
\end{align}

Eq. (\ref{Hrelation}) is the dispersion relation for the growth rate $\sigma(k)$ in terms of wave-numbers $k$. 

Note that for the case of no resistance, $\nu=0$, the dispersion relation reduces to that found by many prior studies on the collective dynamics of micro-swimmer suspensions \cite{Saintillan08a, Subramanian09, Hohenegger10, Subramanian11, Lushi12, Krishnamurthy15, Lushi16, Stenhammar17, Lushi18, Saintillan18, Skultety20}. The difference here is the appearance of $h(\nu)$ in the definition of $a$, and of $(k^2 + \nu^2)/k^2$ in Eq. (\ref{Hrelation}).

\vspace{-0.2in}
\subsection{Small-wave-numbers asymptotic approximation}\label{SmAsyAppr}
\vspace{-0.1in}

The dispersion relation Eq. (\ref{Hrelation}), $\mathcal{F}(\sigma , k) = 0$ cannot be solved exactly for the growth rate $\sigma(k)$. To gain insight into the behavior of the system, we look for long-wave (small wave-number $k$) asymptotic solutions. 

We assume a power series expansion $\sigma = \sigma_0 + \sigma_1  k + \sigma2  k^2 + ...$ for small $k$. The coefficients $\sigma_0, \sigma_1, \sigma_2,...$, can be determined from systematically solving the equations $\mathcal{F}_0(\sigma)=0, \mathcal{F}_1(\sigma)=0, \mathcal{F}_2(\sigma)=0,...$, where these equations arise from a power series expansion in $k$ of the dispersion relation Eq. (\ref{Hrelation}) as in $0=\mathcal{F}(\sigma , k) = \mathcal{F}_0(\sigma) + \mathcal{F}_1(\sigma)  k + \mathcal{F}_2(\sigma) k^2 + ...$. Details of the derivation can be found in \cite{Almoteri23}.

We obtain two distinct asymptotic solutions, or branches, for the growth rate $\sigma$:
\begin{align}
\sigma_{H_1}(k) &=\frac{ (-\alpha \gamma)k^2}{5(k^2+\nu^2)}h(\nu) -  \frac{15(k^2+\nu^2)}{7(-\alpha \gamma)}h(\nu)  -D k^2 +..., \label{Hrelationexpan1_rev}\\
\sigma_{H_2}(k)&=\frac{ (k^2+\nu^2)}{(-\alpha \gamma) }h(\nu)  -Dk^2 +...  .\label{Hrelationexpan2_rev}
\end{align}

For $\nu=0$ we recover the Stokes flow results \cite{Saintillan08a, Saintillan08b} where $\sigma_{H_1}(k=0,\nu=0)=-\alpha \gamma/5$ then decreases with increasing $k$, whereas $\sigma_{H_2}(k=0,\nu=0)=0$ which then increases with small increasing $k$.

Translational diffusion with constant rate $D$ has a stabilizing effect for any type of suspension, as seen in the $\mathcal{O}(k^2)$ terms of both $\sigma_{H_1}(k)$ and $\sigma_{H_2}(k)$.  

{\bf Pullers, $\nu>0$:} We observe that both branches of $\sigma_{H}(k)$ are negative for puller swimmers ($\alpha>0$) of any shape $\gamma$, for all values of resistance $\nu>0$, which means a puller suspension is stable under perturbations. 

{\bf Pushers, $\nu>0$:} For pusher swimmers ($\alpha<0$) however, both $\sigma_{H_1}(k)$ and $\sigma_{H_2}(k)$ can be positive for non-spherical swimmers ($\gamma \neq 0$). It is easy to see that $\sigma_{H_2}(k=0)=0$, and increases with small increasing $k$. Thus pusher suspensions are expected to be unstable under perturbations from the uniform isotropic case.

{\bf Notable:} Here $\lim _{k \rightarrow 0}\sigma_{H_1}(k, \nu>0)=0$ for small $\nu>0$ values, and this is in contrast to the Stokes case where $\sigma_{H1}(k=0, \nu=0)=(-\alpha \gamma)/5$. There is a discontinuity in $\sigma_{H1}(\nu)$ when going from $\nu=0$ to $\nu>0$.  

 \begin{figure}[htpb]
\includegraphics[width=2.4in]{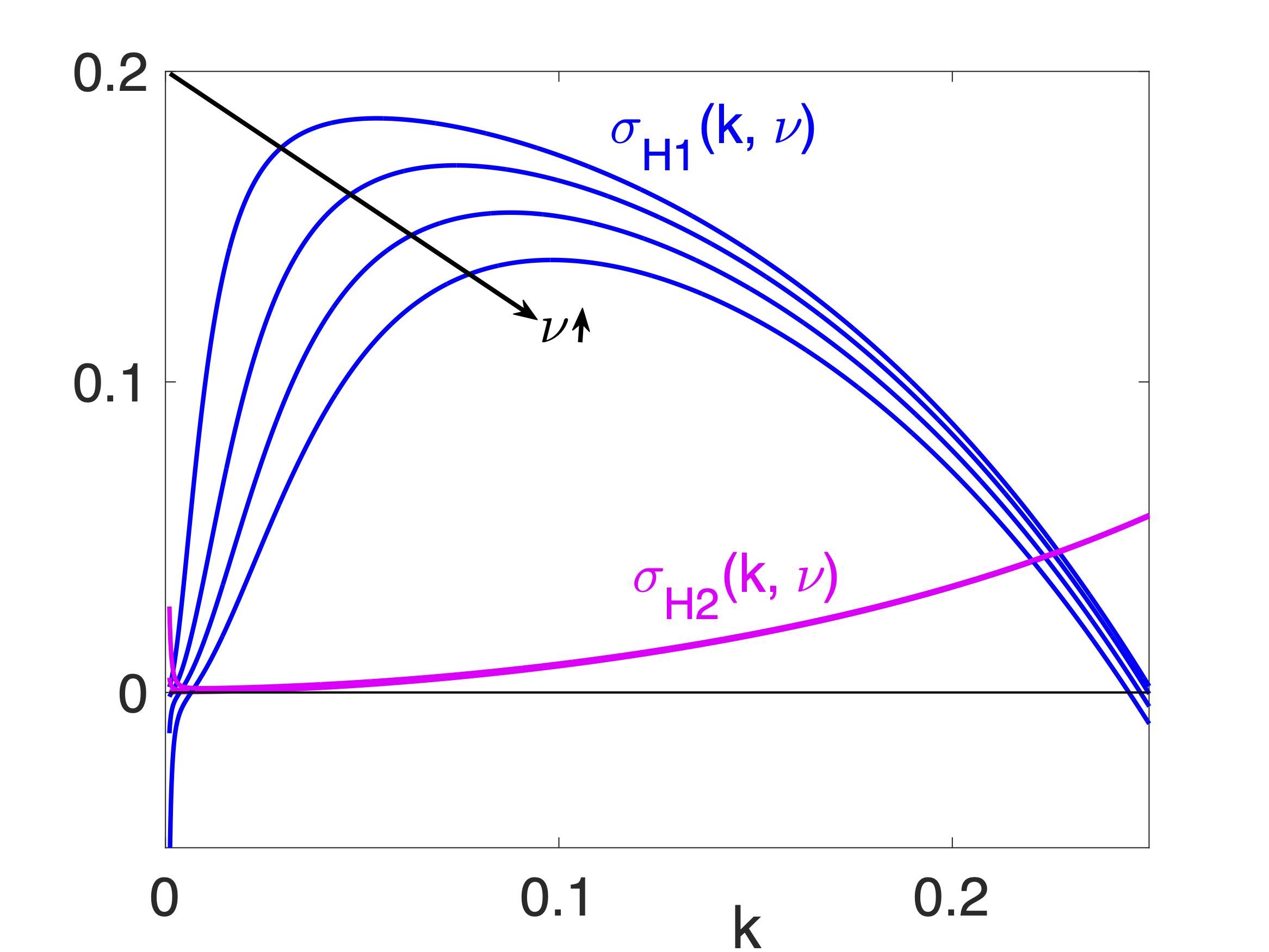}
\vspace{-0.0in}
\caption{Plot of the first three terms from the asymptotic expressions Eqs. (\ref{Hrelationexpan1_rev}, \ref{Hrelationexpan2_rev})  of $\sigma_{H1}(k)$ and $\sigma_{H2}(k)$ for elongated pushers $\alpha=-1, \gamma=1$ with no translational diffusion $D=0$ for values of $\nu=0.01, 0.02, 0.03, 0.04$. 
}
\label{fig:asymptoticplot}
\end{figure}

In Fig. \ref{fig:asymptoticplot} we plot the first three terms from the asymptotic expressions of $\sigma_{H1}(k)$ and $\sigma_{H2}(k)$ from Eqs. (\ref{Hrelationexpan1_rev}, \ref{Hrelationexpan2_rev}) for elongated pusher swimmers $\alpha=-1, \gamma=1$ for a few small but non-zero values of $\nu=$. We see that $\sigma_{H_1}(k\approx 0, \nu \approx0) \approx 0$ and it increases from there for small increasing $k$. Eventually it reaches a maximum and then decreases. Thus the asymptotic results predict a band of wave-numbers for which  $\sigma_{H1}(k)>0$, and this band decreases with increasing $\nu$.

Note that our asymptotic approximations in Eqs. (\ref{Hrelationexpan1_rev}, \ref{Hrelationexpan2_rev}) break down when $k$ is significantly smaller than $\nu$, and this can be seen in Fig.  \ref{fig:asymptoticplot} by the plots of $\sigma_{H1}(k)$ and $\sigma_{H2}(k)$ near $k\approx 0$ for the larger values of $\nu$.

\vspace{-0.1in}
\subsection{Numerical solution of the dispersion relation}\label{sect:numdisprel}
\vspace{-0.1in}

 \begin{figure}[htpb]
\includegraphics[width=3.5in]{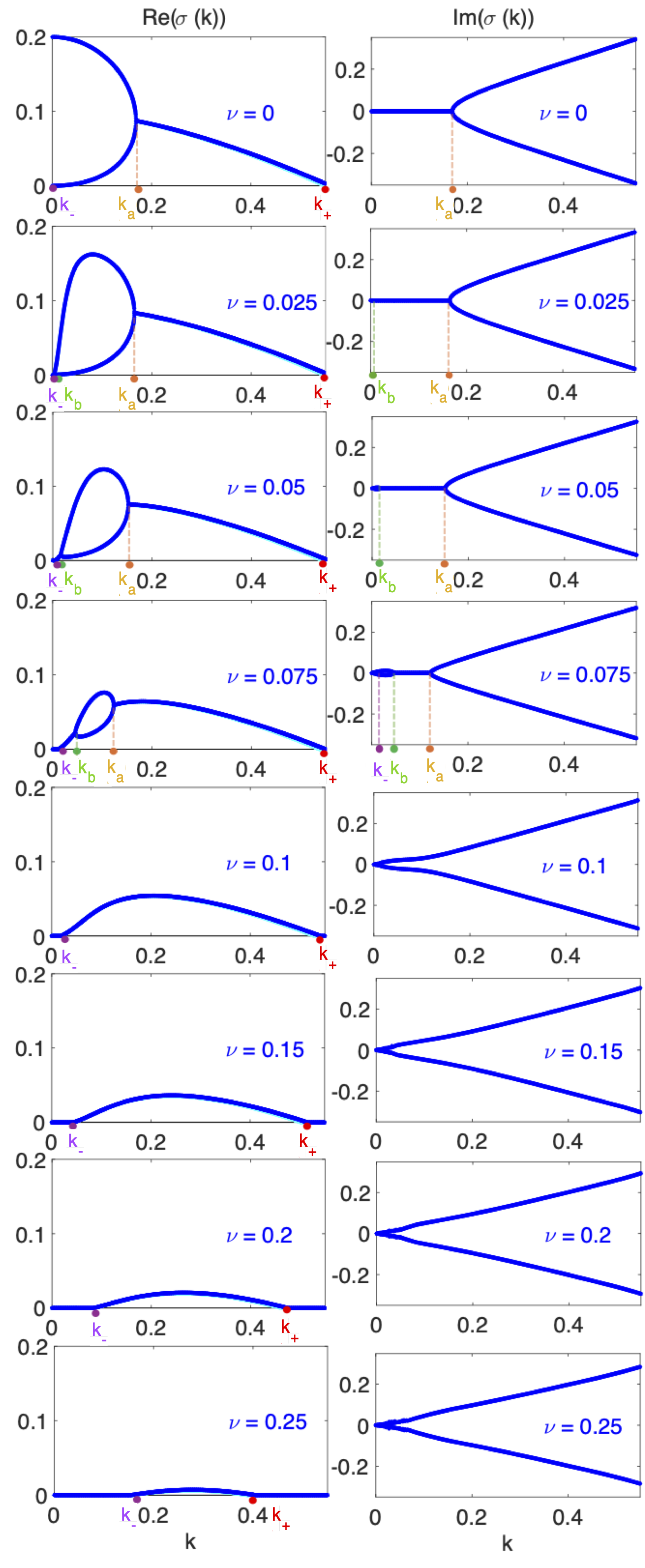}
\vspace{-0.1in}
\caption{Real and imaginary parts of the growth rate $\sigma_{H}(k)$ obtained from numerically solving Eq. (\ref{Hrelation}) for various resistance parameters $\nu= 0, 0.025, ...,0.2$, $D=0$.
Whenever possible, we indicate on the $k$-axis the critical wave-numbers {\color{purple}$k_{-}$},  {\color{green}$k_{b}$},  {\color{orange}$k_{a}$}, and  {\color{blue}$k_{+}$} with purple, green, orange and red circles.
}
\label{fig:SigmaH}
\end{figure}

{\bf Methodology:} We can solve numerically the dispersion relation Eq. (\ref{Hrelation}), $\mathcal{F}(\sigma(k,\nu), k)=0$ for the growth rate $\sigma(k,\nu)$ for each resistance value $\nu$ and wave-numbers $k$ by using an iterative quasi-Newton solver (e.g., Matlab's {\it fsolve} with a trust-region search) \cite{Almoteri23}. We start with $\nu=0$ and then advance for small $\nu$ increments $d\nu$, and for each $\nu$ value we advance in small $k$ increments $dk$ starting from $k=0$. For small $k$ and small $\nu$, we use as initial guess the asymptotic expansions in Eqs. (\ref{Hrelationexpan1_rev}, \ref{Hrelationexpan2_rev}). Further than that, when solving for $\sigma(k+dk,\nu+d\nu)$ we use as an initial guesses linear combinations of $\sigma(k,\nu+d\nu)$ and $\sigma(k+dk,\nu)$ and a solver tolerance of $dk^2$ to be able to capture the solutions. 

{\bf Stable cases:} It is clear from the formulas that translational diffusion has a stabilizing effect on all the suspensions as it lowers the real part of $\sigma_H$ branches by $Dk^2$, and this is the case for any resistance value $\nu$. Puller suspensions ($\alpha>0$) are stable, as are spherical swimmers ($\gamma=0$), for any resistance values $\nu$ according to linear theory, hence we will not discuss them further here. Note however that those suspensions may yet develop instabilities due to other interactions, e.g. steric or aligning \cite{Ezhilan13}.

The case for pusher suspensions ($\alpha>0$) is the most interesting and we will discuss it at length. The numerical solution for the growth rate $\sigma_H(k)$ of elongated pusher swimmers ($\gamma=1$, $\alpha=-1$) with no translational diffusion $D=0$ is shown in Fig. (\ref{fig:SigmaH}) for various resistance values $\nu$. There are two branches of the growth rate $\sigma$, as predicted by the asymptotic analysis in Eqs. (\ref{Hrelationexpan1_rev}, \ref{Hrelationexpan2_rev}) for small $k$. Here we see that for larger $k$ the branches develop an imaginary part.

{\bf Stokes case:} We first summarize the case of homogeneous flow $\nu=0$, a well-known results discussed in many previous studies  \cite{Saintillan08a, Subramanian09, Hohenegger10, Subramanian11, Lushi12,  Krishnamurthy15, Lushi16, Stenhammar17, Lushi18, Saintillan18, Skultety20}, and here shown on the top of Fig. (\ref{fig:SigmaH}). For pushers, $\sigma_{H_1} (k=0)=0.2$ and it decreases with $k$, whereas $\sigma_{H_2}(k=0)=0$ and it increases with $k$. There's a finite range of wave-numbers $k \in (k_{-}(\nu=0)=0, k_{+} \approx 0.57)$ for which $Re(\sigma_H(k)>0$ and $Im(\sigma_H(k)=0$ for both branches, hence a system corresponding to these wave-numbers would see growth of the perturbations from the uniform isotropic state. At $k_a \approx 0.18$, the real parts of the two branches of $\sigma_H$ merge, then decrease together until they cease to be positive at $k_{+} \approx 0.57$. The imaginary parts of $\sigma_H$ branches are both zero for $k<k_a$ and then split and increase in magnitude with increasing $k$.

{\bf Critical wave-numbers:} We have defined $k_{-}(\nu)$ and $k_{+}(\nu)$ as the bounds of the wave-number range $k_{-}(\nu)<k<k_{+}(\nu)$ for which $Re(\sigma_{H}(\nu,k))>0$. We saw in the homogeneous case $k_{-}(\nu=0)=0$ and $k_{+}(\nu=0)\approx 0.57$. 

We define $k_b(\nu)>k_{-}(\nu)$ and $k_a(\nu)<k_{+}(\nu)$ as the bounds of the wave-number range $k_{b}(\nu)<k<k_{a}(\nu)$ where the two branches of $Re(\sigma_H)$ separate and merge again respectively. For homogeneous flows, $k_{a}(\nu=0)\approx 0.18$, but note that a value for $k_{b}(\nu=0)$ does not exist. 

These critical wave-numbers are indicated in the plots of $\sigma_H(k,\nu)$ in Fig. (\ref{fig:SigmaH}). They show that for increasing values of resistance $\nu$, we have:\\
- for $k \in (0,k_{-})$, $Re(\sigma_H)$ is not positive.\\
- for $k \in (k_{-}, k_{b})$, $Re(\sigma_{H1})=Re(\sigma_{H2})>0$, but $Im(\sigma_{H}) \neq 0$, complex conjugate solutions.\\
- for $k \in (k_{b}, k_{a})$, $Re(\sigma_{H})>0$ but have split up, whereas $Im(\sigma_{H}) =0$, real solutions.\\
- for $k \in (k_{a}, k_{+})$, $Re(\sigma_{H1})=Re(\sigma_{H2})>0$, but $Im(\sigma_{H}) \neq 0$, so complex conjugate solutions.\\
- for $k >k_{+})$, $Re(\sigma_H)$ is not positive.

For $k \in (k_{-}, k_{+})$, since $Re(\sigma_{H})>0$, we expect a growth of the perturbations, and when $Im(\sigma_{H}) \neq 0$ we expect an oscillation as well. When $Re(\sigma_H)$ is not positive, any perturbations are expected to decay. 

\begin{figure}[htbp]
\centering
\includegraphics[width=3.3in]{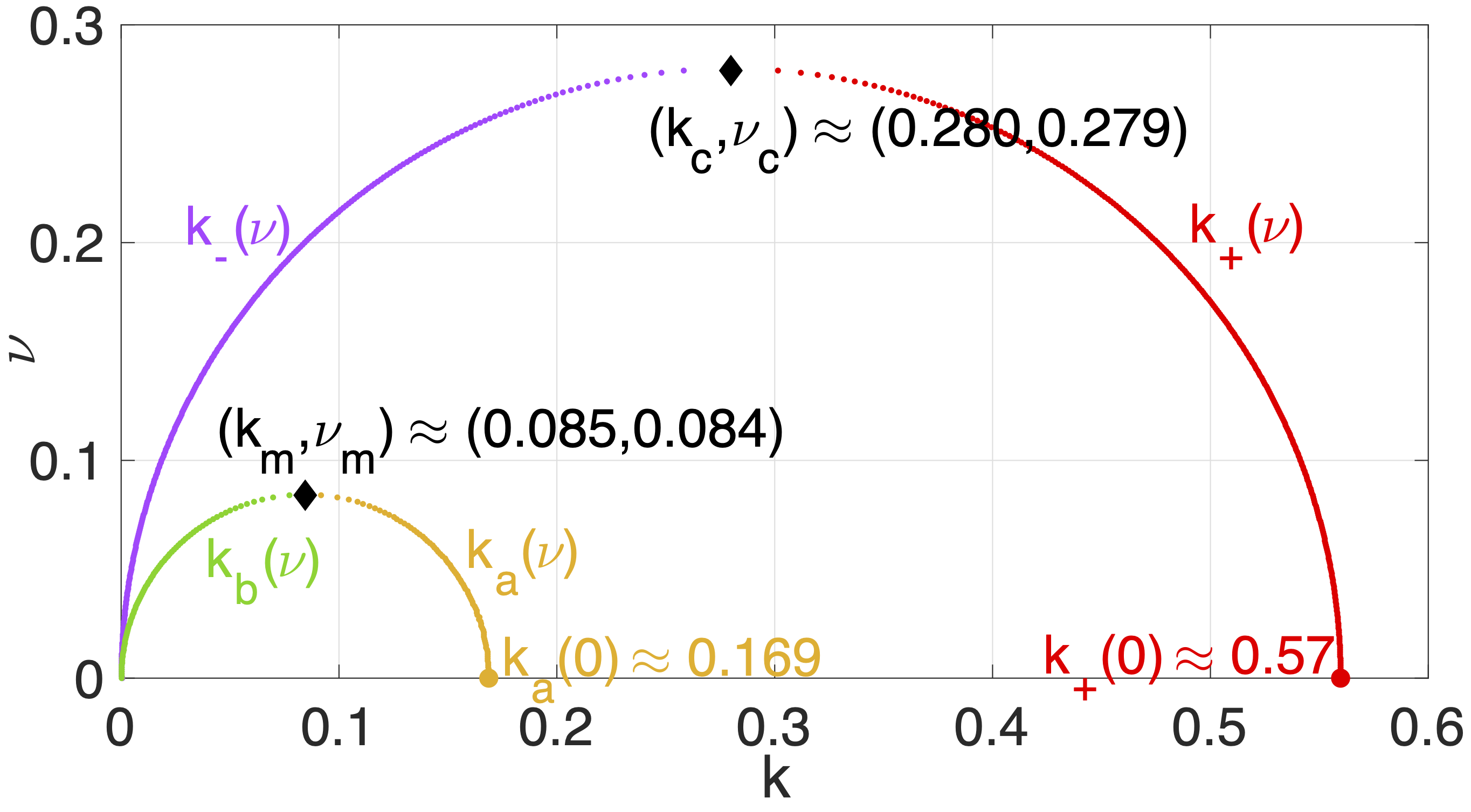}
\vspace{-0.1in}
\caption { Plots of the critical wave-numbers  {\color{purple}$k_{-}$},  {\color{green}$k_{b}$}, {\color{orange}$k_{a}$}, {\color{blue}$k_{+}$} with varying resistance $\nu$ and wave-numbers $k$. 
} \vspace{-0.1in}
\label{fig:knuplot}
\end{figure}

 In Fig. (\ref{fig:knuplot}) we plot $k_{-}, k_b, k_a, k_{+}$ for varying $\nu$. 
 
{\bf Critical resistance $\nu$ values:}  At $\nu_m \approx 0.084$, we have $k_b(\nu) = k_a(\nu)$ and $Re(\sigma_H)$ of the two branches merge and remain merged for $\nu > \nu_m$, whereas their imaginary parts are nonzero and conjugates. 

At $\nu_c \approx 0.279$ we have $k_{-}(\nu) = k_{+}(\nu)$ and $Re(\sigma_H)$ of the two branches are no longer positive parts for any $k$. Hence, linear theory tells us that having $\nu > \nu_c$ turns off the hydrodynamic instability for pusher suspensions for {\it any} wave-numbers $k$, thus {\it any}  system size $L=2 \pi/k$.

\begin{figure}[htbp]
\centering
\includegraphics[width=2in]{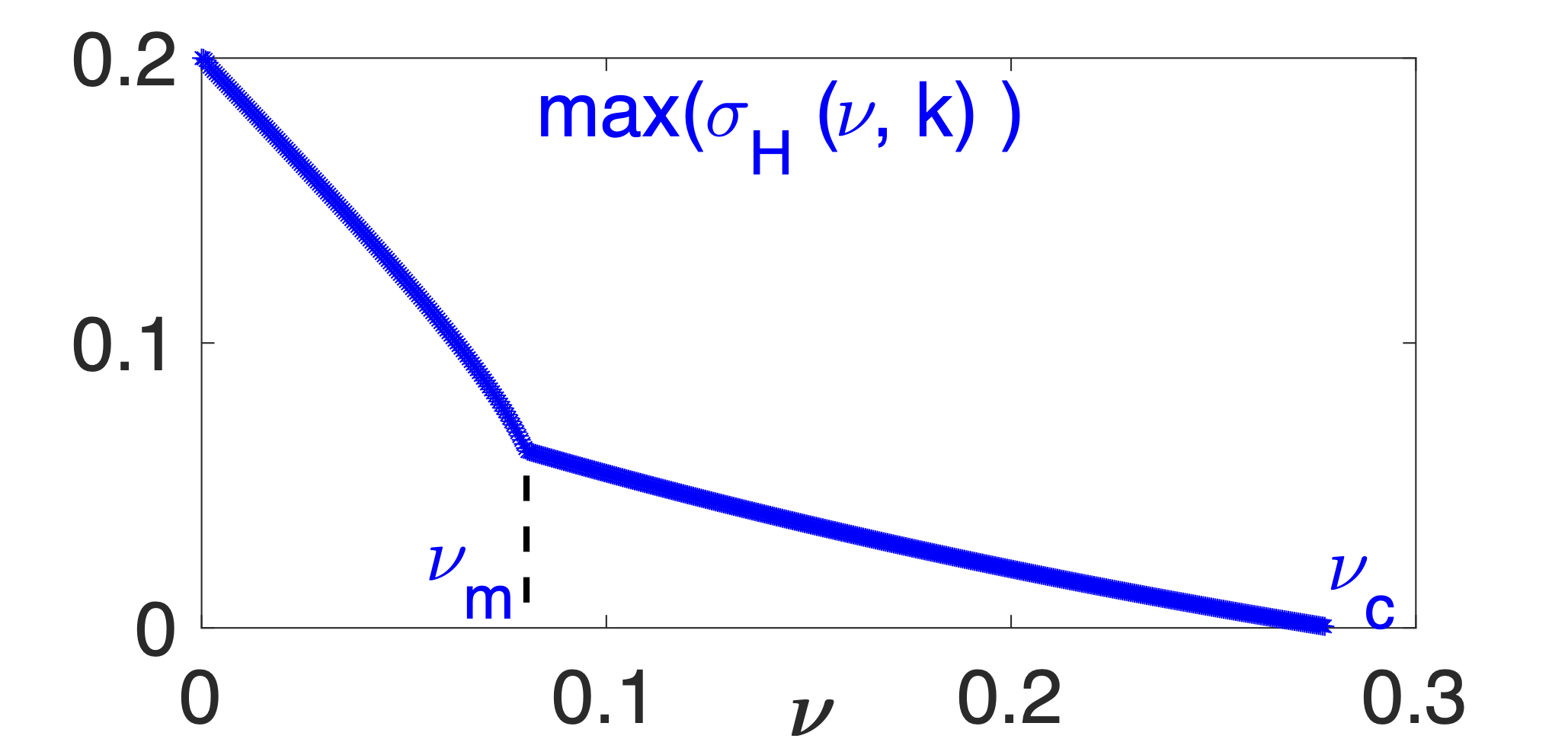}
\vspace{-0.1in}
\caption { Plot of $\max(Re(\sigma_H (\nu)))$ for varying resistance $\nu$. Note the critical values $\nu_m \approx 0.084$ and $\nu_c \approx 0.279$.
}
\label{fig:maxsigmaH}
\end{figure}
\vspace{-0.1in}

Note from Fig. (\ref{fig:SigmaH}) and  the plot of $\max(Re(\sigma_H (\nu)))$ in Fig. (\ref{fig:maxsigmaH}) 
that for any wave-numbers $k$, if $0< \nu_1 < \nu_2$, then $Re(\sigma_H ( k,0) ) > Re(\sigma_H (k, \nu_1) ) > Re(\sigma_H ( k, \nu_2) )$. Therefore fluctuations from the uniform isotropic state in pusher suspensions are less unstable for heterogeneous flows with increasing resistance and also less unstable than homogeneous flows. This agrees with our analysis of the system's configurational entropy.

\vspace{-0.1in}
\section{Nonlinear System's Simulations}
\vspace{-0.1in}
\subsection{Numerical method}
\vspace{-0.1in}

Linear stability analysis suggested parameters for which  perturbations from the uniform isotropic state will grow for given domain sizes, but it does not tell us what the dynamics will look like. To investigate the dynamics of the motile suspensions, we perform numerical simulations of the full nonlinear system, Eqs. (\ref{PsiEqnNondimensional}) and  (\ref{nondimfluid}). 

Since a full 3D system is computationally heavy due to three space and two orientation variables, we focus instead a periodic 2D system in which the particles are constrained to move and rotate in the $(x,y)$-plane with orientation parameterized by one angle $\theta \in [0, 2\pi)$, so that the direction is $\mathbf{p}= (\cos\theta , \sin\theta,0)$. The domain is discretized uniformly with $M=128-256$ points in the $x$ and $y$ directions and $M_{\theta}=16-32$ points in the angle. 

We use second-order accurate finite differences to calculate the flux terms in the conservation equation. The trapezoidal rule is used to compute the integrals in orientation $\theta$, e.g. for the $\Sigma^p$ and $\Phi$. As the computational domain is periodic, we employ spectral methods to solve the fluid equations, as in Eq. (\ref{teldaveloci}). We integrate Eq. (\ref{PsiEqnNondimensional}) using a second order Adams-Bashforth scheme with sufficiently small time-steps to keep the calculations stable. 

The initial condition is chosen to be a random perturbation around the uniform isotropic state \cite{Saintillan08a, Lushi12, Lushi18}:
\begin{align}
\Psi(\mathbf{x},\mathbf{\theta},0) = \frac{1}{2\pi} \left[1+\sum_{i} \epsilon_i \; cos (\mathbf{k}_0 \cdot \mathbf{x} + \xi_i)P_i (\theta) \right], 
\end{align}
where $|\epsilon_i | <0.01$ are randomly-chosen small coefficients, $ \xi_i $ is a random phase, and $P_i (\theta)$ is a third order polynomial of $\sin (\theta)$ and $\cos (\theta)$ with random $O(1)$ coefficients. 

The results presented here are for elongated pusher swimmers $\gamma=1$, $\alpha=-1$. Translational and rotational diffusions are included with coefficients $D = d_r = 0.01$. 

We choose a periodic square box with side $L = 25$ which allows for enough unstable modes according to our linear stability. Specifically, $k=2\pi/L \approx 0.25$ is close  to $k_c \approx 0.28$ in Fig. (\ref{fig:knuplot})) and thus has $Re(\sigma_H(k))>0$ and expected growth of perturbations for a wide range of resistance values $\nu < \nu_c$ according to the plots in Fig. (\ref{fig:SigmaH}). In future work we will examine other box sizes.


\begin{figure}[htpb]
\centering
\includegraphics[width=3.2in]{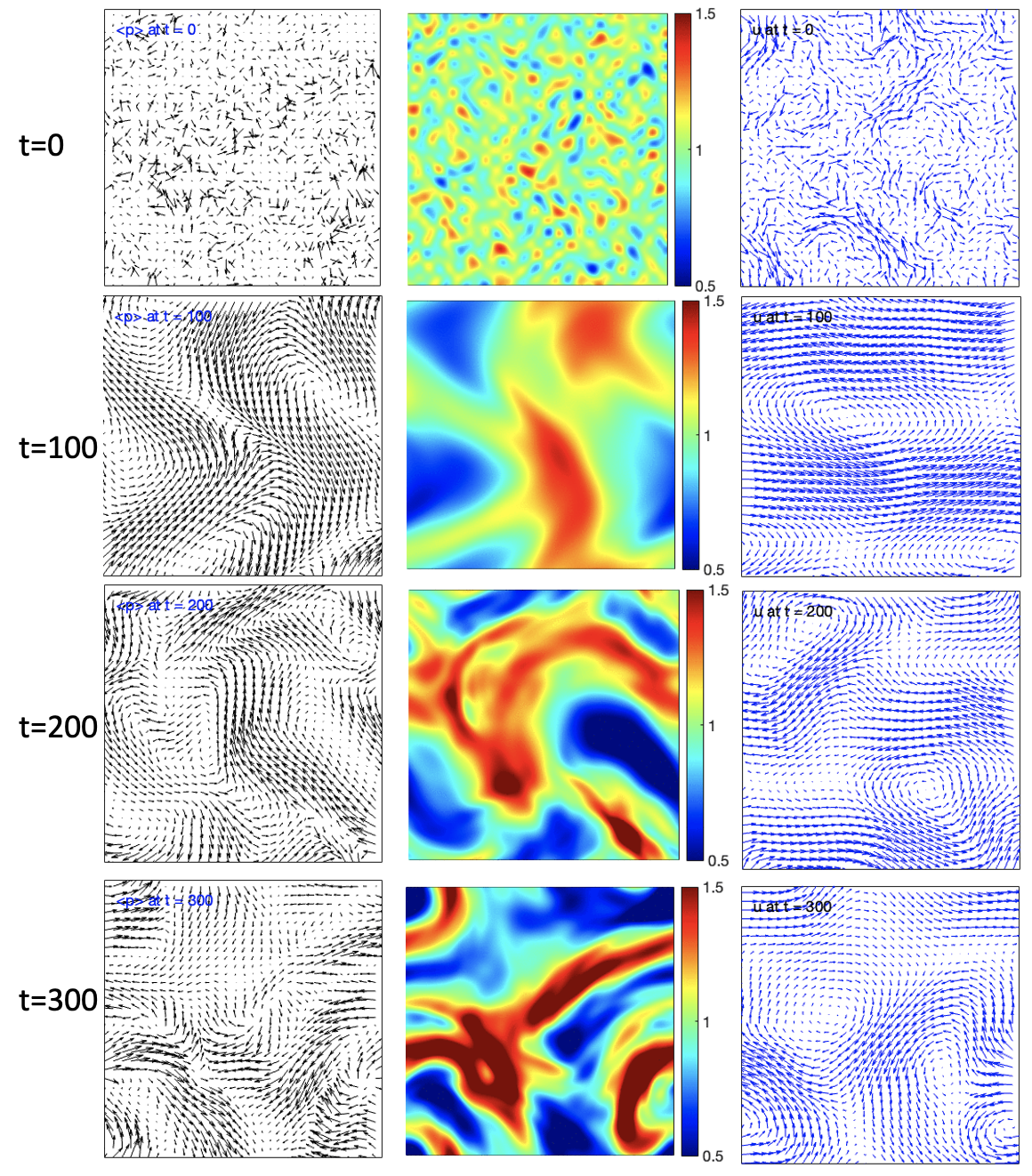}
\vspace{-0.2in}
\caption[The swimmer director, concentration, and  fluid velocity for $\nu=0$.]{The swimmer director $<\mathbf{p}>$, concentration $\Phi$ and fluid velocity $\mathbf{u}$ at times $t=0, 100, 200, 300$ for $\nu=0$. 
}
\label{fig:EffectBrinkman0}
\end{figure}

\begin{figure*}[htpb]
\centering
\includegraphics[width=1.8\columnwidth]{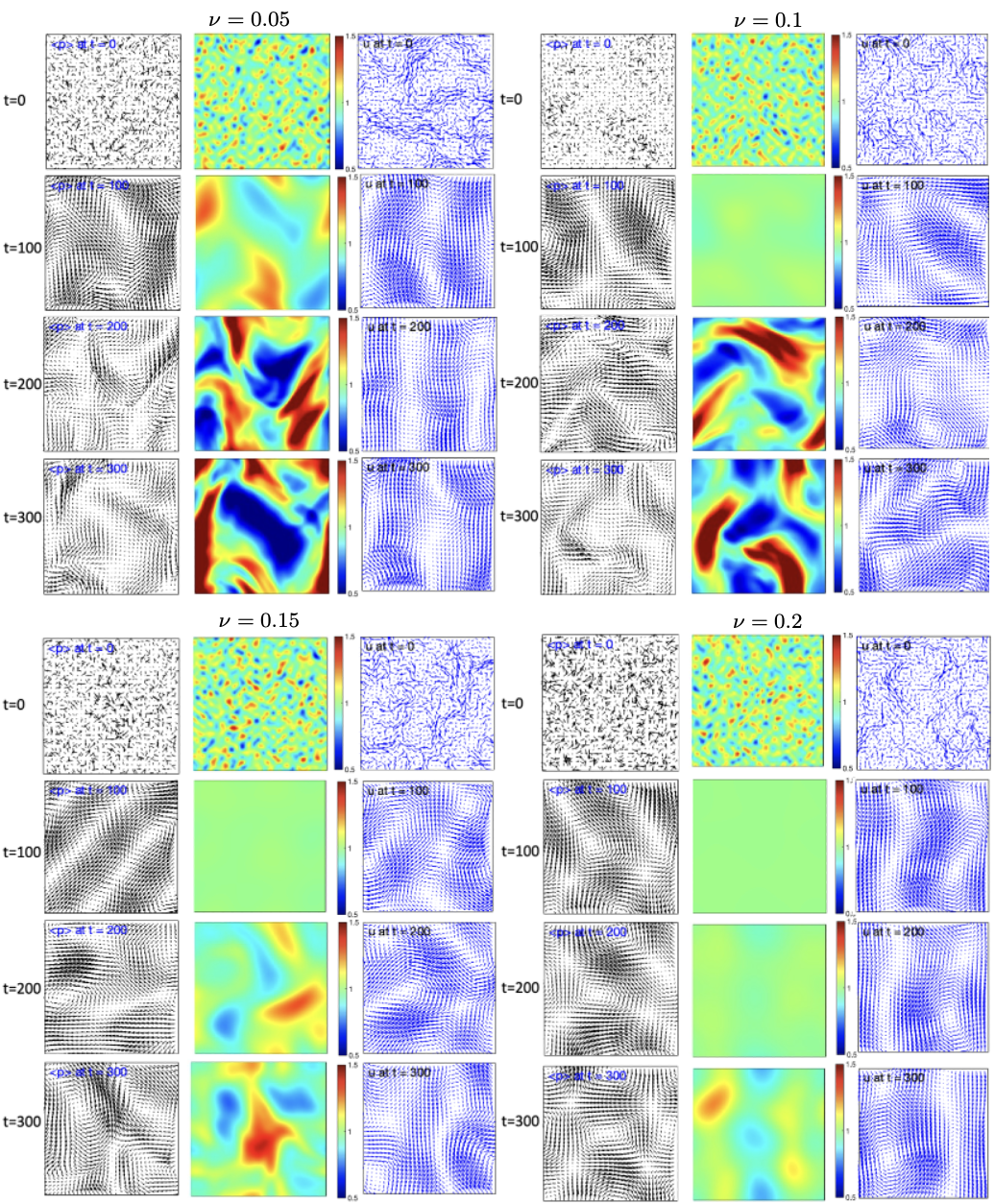}
\vspace{-0.2in}
\caption{The swimmer director $<\mathbf{p}>$, concentration $\Phi$ and fluid velocity $\mathbf{u}$ at times $t=0, 100, 200, 300$ for $\nu=0.05$ (top-left), $\nu=0.1$ (top=right), $\nu=0.15$ (bottom-left), $\nu=0.2$ (bottom=right).
}
\label{fig:EffectBrinkmanNonZero}
\end{figure*}

\vspace{-0.2in}
\subsection{The effect of the Brinkman resistance}
\vspace{-0.2in}

Figs. \ref{fig:EffectBrinkman0} to \ref{fig:EffectBrinkmanNonZero} present simulations of the dynamics of an initially isotropic suspension for various values of the hydrodynamic resistance parameter, $\nu \in 0..0.25$, suggested from the linear stability analysis. 
Select movies are included in the Supplemental Materials.

For $\nu$ zero or small, at short times we obtain the dynamics observed for Stokesian swimmer suspensions \cite{Saintillan08a, Saintillan13}. The fluctuations decay and the concentration field becomes smoother. The mean director and velocity fields quickly become smooth and correlated on scales on the order of the box size. At longer times the concentration field begins to develop strong fluctuations at wavelengths on the order of the box size, whereas the director and velocity fields remain correlated over large scales. The strong fluctuations are dynamic: their magnitude stabilizes due to diffusion but their shape and position keep evolving, with dense concentration bands regions constantly merging, breaking up, and reorganizing. 

For nonzero and increasing $\nu$, e.g. $\nu=0.1$, we notice a delay in the onset of the instability and a decrease in the magnitude of the concentration bands in comparison to the Stokesian case. At $\nu=0.15$, the onset of the instability is further delayed and the concentration bands are smaller, indicating a dampening effect on the fluctuations. At high resistance $\nu_c=0.25$ and further (not shown), the instability is considerably suppressed and the perturbations decay to zero. These results suggest that pushers have difficulties accumulating and moving collectively due to the higher frictional resistance.

Recall that the linear stability analysis yielded the critical resistance value $\nu_c \approx 0.28$ beyond which the hydrodynamic instability would be completely suppressed as $Re(\sigma_H(\nu>\nu_c))<0$. In the nonlinear simulations however we see that the complete suppression of the hydrodynamic instability happens even for $\nu = 0.25<\nu_c$. This can be explained by the presence of other stabilizing terms such as translational and rotational diffusions, as well as the nonlinear coupling between all the terms.

\begin{figure}[htpb]
\centering
\includegraphics[width=3.3in]{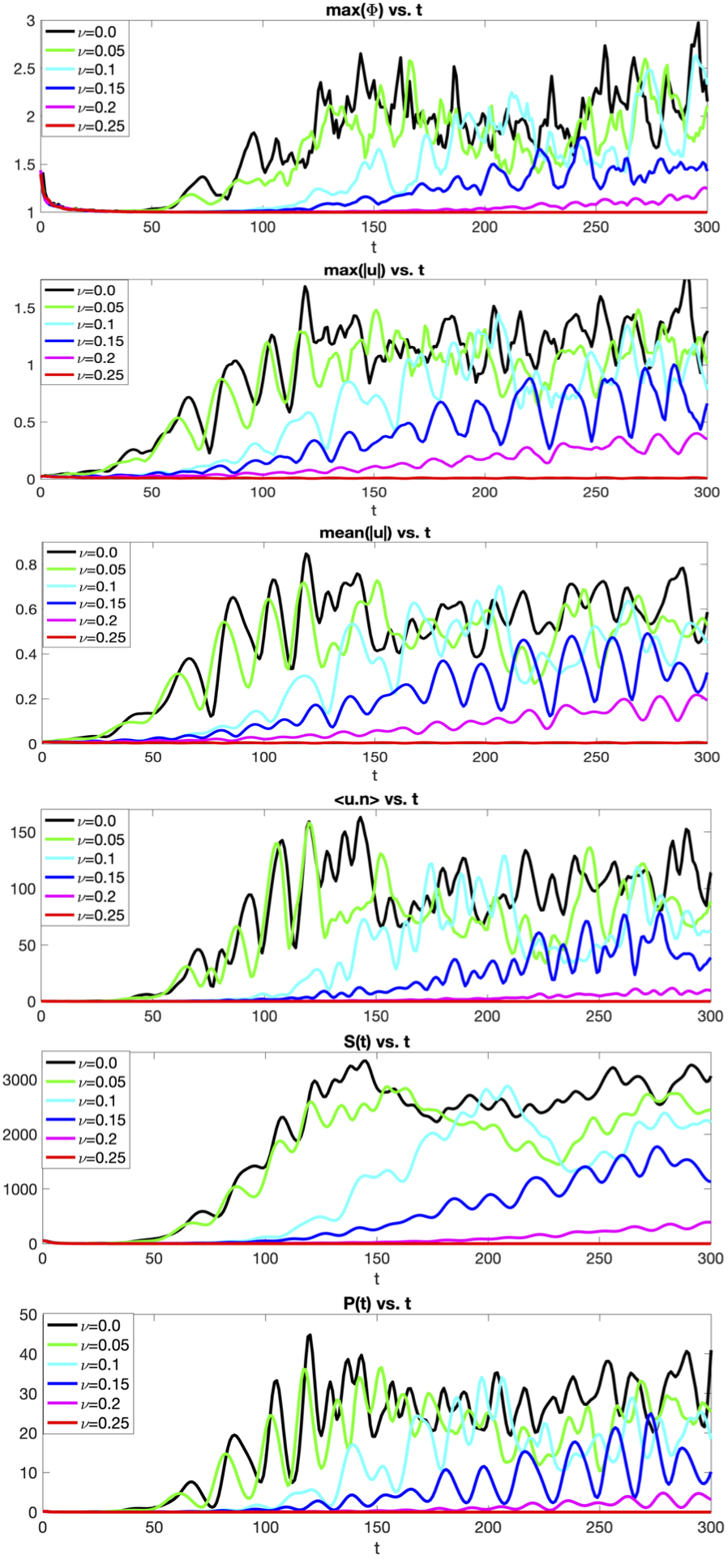}
\vspace{-0.2in}
\caption{ $max(\mathbf{\Phi})$, $mean (|\mathbf{u}|)$, and $max(|\mathbf{u}|)$, $<\mathbf{u} \cdot \mathbf{n}>(t)$, $\mathcal{S}(t)$ and $P(t)$  for different $\nu$.
} \vspace{-0.2in}
\label{fig:comparisons}
\end{figure}

We can quantify the observations from Figs. \ref{fig:EffectBrinkman0} to \ref{fig:EffectBrinkmanNonZero} by tracing the evolution in time of several quantities, shown in Fig. \ref{fig:comparisons}. We focus on the concentration $\Phi$, fluid velocity $\mathbf{u}$, swimmer alignment with the flow $<\mathbf{u} \cdot \mathbf{n}>$, system configurational entropy $\mathcal{S}$ (see Eq. (\ref{EnEq}) ) and the swimmer's global input power $P(t)$ (the right-hand side of Eq. (\ref{EnEq1}) to gain more understanding of how the Brinkman resistance contributes to the system dynamics. 

In particular, Eq. (\ref{EnEq1}) helps us elucidate the mechanisms at play. The left-hand side of the equation has terms representing the rate of viscous dissipation in the fluid and the effect of the viscous resistance, whereas the right-hand side represents the active input power $P(t)$ generated by the immersed particles. A consequence of Eq. (\ref{EnEq1}) is that for pushers ($\alpha<0$) the input power is largest when the particles are aligned with the axes of extension of the rate-of-strain tensor, and this alignment occurs for any particle whose orientation dynamics is governed by Jeffery’s Equation (\ref{pdot}), thus we can expect the input power to grow in time in agreement with the existence of an instability \cite{Saintillan08b, Saintillan13}. For $\nu=0$ the swimmers' global input power $P(t)$ is initially small and increases with time as the hydrodynamic instability due to collective swimming emerges, eventually reaching a plateau when the system enters a state of statistical equilibrium. When Brinkman resistance is introduced, the swimmers' global input power decreases, indicating that the system requires less power to maintain the fluid flow and dynamics. The Brinkman resistance to the flow of the fluid reduces the velocity and, hence, the kinetic energy of the fluid. The effect becomes more pronounced when near $\nu_c$, and the global input power quickly decays to zero.

The observed effects can be further explained by considering the competition between the tendency of pusher swimmers to accumulate and form concentration bands and the dampening effect of the fluid flow resistance. As the resistance $\nu$ is increased, the frictional forces acting on the fluid and bacteria become stronger, leading to a decrease in the amplitude of the concentration bands and a delay in the onset of the instability. Moreover, at higher values of $\nu$, the damping effect dominates over the tendency of bacteria to form concentration bands, resulting in the suppression of the hydrodynamic instability and a decrease in the collective motion of bacteria.

The nonlinear simulations show a consistent pattern that is in agreement with the entropy and linear stability analysis: resistance has a dampening effect on the pusher suspension dynamics, indicating its ability to hinder the onset and development of the collective swimming instabilities, and to completely suppress it if sufficiently large.

\vspace{-0.1in}
\section{A Note on the Parameters}
\vspace{-0.2in}

We include below the dimensional values for the parameters and critical values to help the readers connect the predictions here to experiments. 

{\it Bacillus subtilis} bacteria have length $\ell=4 \mu m$, width $2\rho=1 \mu m$, speed $U=20 \mu m/s$. Following our single swimmer model's Eq. (\ref{xdotoneswimmer}) where $U=f^f/(2\xi)$ with $\xi=6 \pi \mu \rho$ where the fluid viscosity is $\mu = 10^{-3}pN s /\mu m^{2}$, we get a dimensional flagellar strength $f^f = 12 \pi \mu \rho U = 12 \pi \times 10^{-2} pN \approx 0.38 pN $, which is close to experimental measurements of  $0.43 pN$ \cite{Drescher11}. 

Assuming a mean swimmer cell concentration of $n=0.5-2 \times 10^{10} cells/cm^3$, similar to scenarios in experiments that studied the collective motion of {\it Bacillus subtilis} \cite{Dombrowski04, Tuval05, Aranson07, Sokolov07, Sokolov09, Cisneros11, Sokolov12}, we obtain length-scale $\ell_c=(n\ell^2)^{-1} \approx 3-12.5 \mu m $. The simulation box size corresponds to $25 \ell_c = 78-312 \mu m$, consistent with the aforementioned experiments. 

The critical wave-number $k_c\approx 0.57$ gives a critical system $34.5-137.8 \mu m$ for the instability to develop in homogenous flows ($\nu=0$) for these cell concentrations.  The critical system sizes for specific $\nu>0$ can be calculated using the linear analysis plots like Figs. \ref{fig:SigmaH}.

The resulting non-dimensional dipole strength is $|\alpha| = |\sigma_0|/(U \mu \ell_c^2)=24 \pi/ \ell_c^2 = 0.48-7.72$, an $O(1)$ constant. The rotational diffusion of $0.05 rad^2 /s$ from experiments \cite{Drescher11} correspond to non-dimensional $d_r = 7.8-31.3 \times 10^{-3}$.

As $\nu=\ell_c/ \sqrt{K_D}$, the critical resistance value of $\nu_c \approx 0.279$ gives a critical permeability constant $K_{DC}  = 125.5-2007 \mu m^2$. Any permeability $K_D <K_{DC} $ would suppress the swimmer-induced hydrodynamic instability. 

If the porous medium is a dilute suspension of identical solid spheres of radii $\rho_S$, with number density $n_S$ and total volume fraction $\phi_S$, then we can assume the force per unit volume due to permeability (calculated from Darcy's law $\nabla q = \mu/K_{D0} \mathbf{u}$) must be equal to $n_S$ multiplied for the Stokes drag \cite{Vanni00}. This results in a permeability $K_{D0} = ( 2 \rho_S^2)/)(9 \phi_S) $, an approximation that we can use to obtain some estimates. For example, for obstacle spheres of radii $1\mu m$ and {\it Bacillus subtilis} bacteria with density $n=10^{10} cells/cm^3$, the critical solid sphere volume fraction is $0.4\%$, and our theory predicts that any solid sphere volume fraction higher than that should suppress the swimmer-induced instability. 

These values give screening length $25-45\mu m$ and intra-obstacle distance $>$ $8\mu m$, both larger than the typical swimmer size $4 \mu m$. For distances less than the screening length, the fluid velocity is essentially Stokesian \cite{Leshansky09}. 

Note that Brinkman media can also be appropriate representations of dilute gels, e.g. gastric mucus which has density $3-5\%$, when the swimmer
does not directly contact the gel and does not deform the gel network \cite{Mirbagheri16}.

This coarse-grained model considers only dilute swimmer and sparse obstacle suspensions, and it does not include direct swimmer-swimmer and swimmer-obstacle interactions which would significantly impact the overall dynamics. Thus we add the caveat that, as with most theoretical models that cannot possibly include all the complexities in natural and experimental settings, the critical values derived {\em for this model} should be viewed as qualitative and not quantitative predictors, but hopefully still motivate new experimental studies.

\vspace{-0.3in}
\section{Summary and Discussion}
\vspace{-0.2in}

We derived and used a continuum model to study the dynamics of a dilute suspension of micro-swimmers in Brinkman fluid flows. The model consists of a conservation equation for the swimmer positions and orientations, coupled to the fluid flow equations in which the effect of the swimmers' motion is represented by an active particle stress tensor, as done in precursor models \cite{Simha02b, Saintillan08a, Baskaran09, Subramanian09, Skultety20}. We include in the fluid equations a linear resistance or friction term to account for the effect of the medium inhomogeneity or ``porosity''. 

We analyzed the stability of perturbations from the uniform isotropic state of swimmers suspensions. The exact result of the dynamics of the configurational entropy clearly showed that the Brinkman resistance has a stabilizing effect in pusher and puller suspensions alike. In pusher suspensions ($\alpha<0$), the hydrodynamic or Brinkman resistance and the diffusion processes balance and help to drive down the increase in fluctuations. 

The analysis of the linear stability of the isotropic suspensions lead to curious results but the same conclusion: resistance has an overall a stabilizing effect of the instability that results from the hydrodynamic interactions between the swimmers. Whereas in Stokes flow perturbations in pusher suspensions are unstable for a finite range of wave-numbers $k \in \left[ 0, k_{+}(\nu = 0) \approx 0.57\right]$ as a result of hydrodynamical interactions, a result meticulously studied by many colleagues \cite{Saintillan08a, Saintillan08b, Subramanian09, Hohenegger10, Subramanian11}, in Brinkman flows with $\nu \neq 0$, the pusher suspensions are unstable for a smaller range of non-zero wave-numbers   $k \in \left[ k_{-}(\nu), k_{+}(\nu) \right]$  and for resistance values $0 < \nu \leq \nu_c \approx 0.279$. Here $\nu_c$ is the critical resistance value that linear theory suggests is the lower bound for any resistance to completely turn off the instabilities at {\em any} wave-numbers, diffusion constants and swimmer aspect ratio in for pusher suspensions ($\alpha<0$). 

Moreover, we see some peculiar behavior by numerically solving the dispersion relation for the growth rate that results from the linear stability analysis, Fig. \ref{fig:SigmaH}. For Stokes flow, the shear active stresses are unstable at long wavelengths for the case of pushers ($\alpha$). From the growth rate plots, low wave-numbers fluctuations are expected to amplify exponentially at short times, intermediate wave-numbers fluctuations amplify at a lower rate and may exhibit oscillations, whereas high wave-numbers fluctuations decay while also exhibiting fluctuations. We expect the dominant wave-numbers to be the one where $Re(\sigma_H(k))$ is maximized for the given resistance value $\nu$, and these can be found in the Figs. \ref{fig:SigmaH} and \ref{fig:maxsigmaH}.

Full simulations of the system enable us to see the long-time dynamics of pusher suspensions in the parameter regimes indicated by the linear stability analysis. Simulations show the emergence of strong concentration fluctuations that are dynamically unstable--they break up and merge again in quasi-periodic fashion--however, resistance delays the onset and development of the collective swimming and weakens the concentration fluctuations. Resistance thus dampens this hydrodynamic instability presented by collective swimming, and can suppress it altogether if it is sufficiently large.

A notable result of this work, obtained from both the entropy analysis and the linear stability analysis of the uniform isotropic state and confirmed from the nonlinear simulations, is that the Brinkman resistance primarily affects the system dynamics through the bulk-fluid effect on the suspension and not through the decrease in the individual swimmer speed $U_B(\nu)$. This speed $U_B$ does not show at all in the exact equation for the time-evolution of the configurational entropy, has a minor effect in the critical resistance and wave-numbers found in the linear stability analysis (e.g. its inclusion changes $\nu_c$ from $0.269$ to $0.279$), and it has an unnoticeable effect in the nonlinear dynamics observed through simulations. This insight can motivate future studies on adequate approximations in models of swimmer suspensions in complex flows.

We are not aware of any experiments that consider the scenario in our theoretical paper, but hope that this will inspire new work and tests. The setups of the experiments with bead-like gels \cite{Bhattacharjee19a, Bhattacharjee22, MartinezCalvo22, Moore-Ott22}, quasi-2D disordered environments \cite{Makarchuk19, Dehkharghani23}, or colloidal suspensions \cite{Kamdar22}, have micro-structures of a scale larger than can be modeled with the Brinkman approximation used here, but are promising developments on this topic.

Lastly, our current model is limited to dilute non-chemotactic suspensions, but it is possible to add chemotactic interactions \cite{Lushi12, Lushi16, Lushi18, Almoteri25} or an aligning effect to account for higher densities \cite{Ezhilan13}. The system can be adapted to study non-constant and space-dependent friction, creating the effect of soft ``obstacles", and can be used to study transport of micro-swimmers or active matter in porous media or patterned surfaces, as in \cite{Volpe11,Stoop19, Thijssen21}. The combined new simulations and experiments can facilitate the development of optimized control of active flows.

{\bf Acknowledgements:}
The authors gratefully acknowledge support from the fellowships from the Kingdom of Saudi Arabia (Y.A.) and the Simons Foundation grant no. 639018 (E.L). We thank T. Bhattacharjee, S. Datta, C. Hohenegger, F. Guzman-Lastra, A. Morozov, D. Pushkin, D. Saintillan, M. Shelley, and Y. Young for helpful discussions. 


\bibliography{references}

\end{document}